# Isotope effect in tunnelling ionization of neutral hydrogen molecules

X. Wang, [1,2] H. Xu, [1,*] A. Atia-Tul-Noor, [1] B. T. Hu, [2] D. Kielpinski, [1,3] R. T. Sang, [1,3] and I. V. Litvinyuk [1,†]

[1] Centre for Quantum Dynamics and Australian Attosecond Science Facility,
Griffith University, Nathan, QLD 4111, Australia
[2] School of Nuclear Science & Technology, Lanzhou University, Lanzhou, 730000, China and
[3] ARC Centre of Excellence for Coherent X-Ray Science,
Griffith University, Nanthan, QLD 4111, Australia

**Abstract**

It has been recently predicted theoretically that due to nuclear motion light and heavy hydrogen molecules exposed to strong electric field should exhibit substantially different tunneling ionization rates (O.I. Tolstikhin, H.J. Worner and T. Morishita, Phys. Rev. A **87**, 041401(R) (2013) [1]). We studied that isotope effect experimentally by measuring relative ionization yields for each species in a mixed $H_2/D_2$ gas jet interacting with intense femtosecond laser pulses. In a reaction microscope apparatus we detected ionic fragments from all contributing channels (single ionization, dissociation, and sequential double ionization) and determined the ratio of total single ionization yields for $H_2$ and $D_2$. The measured ratio agrees quantitatively with the prediction of the generalized weak-field asymptotic theory in an apparent failure of the frozen-nuclei approximation.

.

The tunnelling ionization of atoms and molecules by intense laser field is a fundamentally important process in the laser-matter interactions. Quantitatively accurate *ab initio* theoretical modelling of this process so far has only been demonstrated for atomic hydrogen [2]. In all other cases the modelling involves a number of approximations, i.e. strong-field approximation (SFA), single-active-electron (SAE) approximation, quasi-static approximation (QSA), frozen-nuclei approximation (FNA), Born-Oppenheimer approximation (BOA), and so on. In particular, for modelling strong-field ionization of molecules the frozen-nuclei approximation is very commonly used. This approximation treats nuclei as classical point particles and models electronic structure of molecules only for specific ("frozen") configurations of nuclei, most often corresponding to equilibrium geometry of neutral molecules. Despite its very wide acceptance, the quantitative validity of FNA was not vigorously tested and its domain of applicability in context of strong-field physics is not clearly defined. For instance, according to FNA ionization rates of mass-isotopes of molecules should be the same, since those isotopes have identical equilibrium geometry and electronic structure in their ground electronic state. Based on that assumption of identical ionization rates the measured ratios of high harmonic generation (HHG) yields for $H_2/D_2$ and $CH_4/CD_4$ isotopic pairs were interpreted as entirely due to nuclear motion between ionization and recollision [3].

Recently, using the generalized weak-field asymptotic theory (WFAT) [4], modified to account for motion of nuclei, Tolstikihin, Worner and Morishita (TWM) theoretically investigated the effect of nuclear mass on the tunnelling ionization rate of hydrogen molecule in a DC electric field [1]. The TWM theory predicts that the nuclear motion results in an increased tunnelling ionization rate in comparison to the frozen-nuclei model, and this increase is more significant for lighter nuclei and weaker laser field. In particular, it predicts that the single ionization rate for $H_2$ is about 1.3 times higher than that of $D_2$ driven by a DC electric field with field strength of F = 0.04 a.u., and the ratio will decrease asymptotically to unity with increasing field strength. The two isotopes have exactly the same electronic states and potential energy curves, the only difference is the nuclear mass, which can lead to different nuclear wavefunctions and different nuclear dynamics in the ionization process. While nuclear mass effects have been observed in many strong-field experiments, such as the multiphoton ionization [5], HHG [3,6], the laser-induced coherent nuclear motion [7], CEP-dependent asymmetry of dissociation [8], and rescattering double ionization [9,10], so far there was no experimental verification of isotope effect on total molecular ionization rate as predicted by TWM. Here we verify this prediction by comparing single ionization yields of $H_2$ and $D_2$ under identical conditions.

In order to make a quantitative comparison of ionization rates for $H_2$ and $D_2$, a 50:50 mixed-gas jet is used in our experiment, thus ensuring that $H_2$ and $D_2$ are ionized by laser fields with exactly the same laser parameters - peak intensity, focusing geometry, pulse duration, and with same fluctuations of those parameters during the measurements. Additionally, mixed-gas jet scheme also guarantees that the target densities of $H_2$ and $D_2$ in the interaction region are also identical. That assumption neglects the difference in expansion of the two components of the mixed jet and it was found to be valid by comparing the ionization yields at intensities above saturation for both isotopes. Circularly polarized pulses with duration of 28 fs are used in our experiment. The circular polarization was chosen for several reasons. Firstly, unlike the linear polarized field, whose instantaneous field strength oscillates rapidly in time, the circular polarized pulse has slowly varying field strength, which is closer to the DC field used in the TWM theory. Secondly, circular polarization minimizes possible effect of molecular alignment during the pulse, which was also checked by comparing angular distributions for proton and deuteron dissociation fragments. Thirdly, and most importantly, circular polarization supresses recollision-induced production of high-energy protons and deuterons [10], whose kinetic energy is overlapped with the double ionization channel. Since we only detect charged particles, the dissociation and double ionization contribute differently to the total proton/deuteron count. It is therefore important for proper counting to separate the dissociation/double ionization channels, which are distinguished by their kinetic energy for circular polarization, and would be overlapping for linear polarization. In the experiment, we measured the ratio of total single ionization yields, while the theory calculates the ratio of ionization rates. The isotope effect is most pronounced at low intensities where ionization probabilities are much below unity and the ratios of yields and rates are the same. As the intensity reaches the saturation intensity and ionization probability approaches unity so does necessarily the ratio of yields, which also coincides with the theoretical prediction for the rates becoming the same at high intensity.

The schematic diagram of our experimental setup is shown in figure 1(c) and its detailed description can be found in [11]. Briefly, measurements are conducted in a reaction microscopes [12] (REMI) with 28 fs, 800 nm circular polarized laser pluses from the laser system FEMTOPOWER Compact PRO (Femtolasers). The laser beam is directed into the main chamber of REMI, where it is focused onto a low-density molecular beam by a spherical mirror (f = 75 mm). The molecular beam is generated by a supersonic 2 stage gas jet system, and its width in the REMI chamber along the laser propagation direction is restricted by a slit to a size (~50 μm) less than the Rayleigh length (~250 μm) to limit intensity variation over the interaction region. The laser intensity is controlled by an adjustable iris located before the beam entering the chamber. The laser peak intensities were precisely calibrated *in situ* by using the recoil-ion momentum imaging method [13] and the overall errors were estimated to be 10% for all peak intensities. The ion fragments are detected by a multi-hit delay-line anode detector (RoentDek), and the full 3D momentum of each fragment can be retrieved. The measured momentum spectra are presented in Fig. 1(a) for $D^+$ and $D_2^+$ (inset), and Fig. 1(b) for H + and $H_2^+$ (inset).

After first tunnelling ionization of neutral $H_2$ and $D_2$, the $H_2^+$ and $D_2^+$ can be further dissociated or ionized by the laser pulse. Figure 2 shows the three possible fragmentation channels of $H_2$ (same for $D_2$): (1) single ionization of neutral $H_2$ only ($H_2 \rightarrow H_2^+$) - in this case only $H_2^+$ can be detected; (2) single ionization followed by dissociation ($H_2 \rightarrow H_2^+ \xrightarrow{diss.} H + H^+$) - one proton can be detected because REMI can only detect charged particles; (3) single ionization followed by sequential double ionization ($H_2 \rightarrow H_2^+ \rightarrow H_2^{++} \xrightarrow{CE} H^+ + H^+$) - a pair of protons can be detected. The ion counts from channel 1 and 2 equals to the counts of single ionization because each ion corresponds to one single ionization event. However, for channel 3, as each single ionization event can lead to a pair of protons, the number of single ionization events should be half of the proton counts. To calculate the total number $N_{tot}$ of single ionization events, one has to take all these three channels into account: $N_{tot} = N_1 + N_2 + 0.5N_3$, where $N_i (i = 1,2,3)$ is the ion counts from three channels respectively.

Figure 3(a) presents a typical time-of-flight (TOF) spectrum of the ions detected in the experiment with the laser peak intensity $4.2 \pm 0.4 \times 10^{14} W/cm^2$. Three distinct bands in the TOF spectrum are for $H^+$, $D^+/H_2^+$ and $D_2^+$ peaked around 650 ns, 900 ns, and 1300 ns, respectively. The narrow $H_2^+$ peak overlaps with the broad $D^+$ band in TOF spectrum due to their identical mass-to-charge ratios. We note that channel 2 and 3 cannot be well resolved in TOF spectrum, because from TOF signal only time information, i.e., one dimension of momentum, can be retrieved, while in case of circular polarization, the ion momentum has a "doughnut" like structure in laser polarization plane, as shown in Figs. 1(a) and (b). Therefore, the TOF spectra of channels 2 and 3 are overlapped with each other. Fortunately, REMI is capable of measuring full 3D momentum and determining kinetic energy (KE) of each charged fragment. The measured KE spectra of $H^+$ and $D^+$ from channels 2 and 3 are shown in figure 2(b). Ions from channel 2 have a smaller KE than those from channel 3, making it possible to separate the channels and to compare overall total single ionization yields for the two isotopes. Here we neglect any possible difference in detection probability of protons and deuterons, as well as any dependence of that detection probability on kinetic energy of fragments.

The corresponding KE spectra of $H^+$ and $D^+$ are shown in Fig. 3(b). Two distinct energy bands can be well resolved for both isotopes. The low energy band (KE < 1.5 eV) corresponds to channel 2, which contains bond softening (BS [14]) and the net-two-photon above threshold dissociation (ATD) [15] processes. The high-energy band (KE > 1.5 eV) arises from the enhanced ionization (EI) [16]. The re-collision induced dissociation and non-sequential double ionization channels [9,10] are eliminated by using circularly polarized laser pulses. The molecular ions $H_2^+$ and $D_2^+$ from channel 1 have a narrow energy spread (< 0.1 eV) in near-zero KE region (not shown here), due to the cold supersonic gas jet used in our experiment. Thus, channel 1 can be clearly separated from the KE spectrum for $H_2$.

The ratio of the measured first ionization yields for $H_2$ and $D_2$ as a function of the peak laser field is presented in Fig. 4(a). The experimental yield ratio is in good agreement with the ratio of ionization rates (Fig. 4(a), blue curve) from TWM theory. In the TWM theory the ratio of the ionization rates of $H_2$ and $D_2$ is given by [1] $\Gamma(H_2)/\Gamma(D_2) \approx \exp\left[\left(1 - \frac{1}{\sqrt{2}}\right) 0.0017/F^2\right]$, where F is the field strength in atomic units. The WFAT is valid for a weak field $F \ll F_c$, where $F_c$ is a critical field strength corresponding to complete suppression of potential barrier [4]. In our experiment the peak electric field varies from 0.045 a.u. to 0.10 a.u., well below $F_c$.

Our experimental results are in good agreement (within the error bars) with the TWM theory prediction over the full range of intensities. At the lowest laser field strength of 0.045 a.u., the measured ratio of $H_2^+/D_2^+$ total ionization yields is approximately 1.25, which agrees with the theoretical value of 1.3. With increasing field strength, the measured ratio decrease asymptotically towards unity as predicted by the theory. It is worth noting that for extreme strong field strength the comparison is not reliable, because the ionization probability is saturated and the yield ratio no longer reflects the ionization rate ratio. A variant of the ADK theory [17] is used to estimate the ionization probability for different intensities, and the result is shown in figure 4(b). When F is between 0.045 a.u. and 0.08 a.u., the ionization probability is well below unity (between $6 \times 10^{-4}$ and 0.5), suggesting that the measured yield ratio reflects the ionization rate ratio. For field strengths F > 0.08, the ionization probability starts to saturate, and the yield ratio is mainly given by the gas density ratio.

In conclusion, we experimentally investigated the isotope effect in the photoionization of neutral hydrogen molecules by measuring the ratio of photoionzaiton yields of hydrogen and deuterium in the tunnelling ionization region with circularly polarized infrared laser pulses. Our experimental results are in quantitative agreement with the theory prediction of Tolstikhin, Worner and Morishita of weak-field asymptotic theory in low laser intensity region. We find that at low intensity corresponding to the field strength of 0.045 a.u. light hydrogen molecule is 25% more likely to ionize than its heavy isotope. This observation exemplifies a failure of the frozen nuclei approximation to accurately model single ionization of neutral molecules.

This work was supported by ARC Discovery Project (DP11010101894). X. W. is grateful for support by the scholarship under the State Scholarship Fund via the China Scholarship Council (CSC, File No. 201306180076). H.X. is supported by ARC Discovery Early Career Researcher Award (DE130101628).

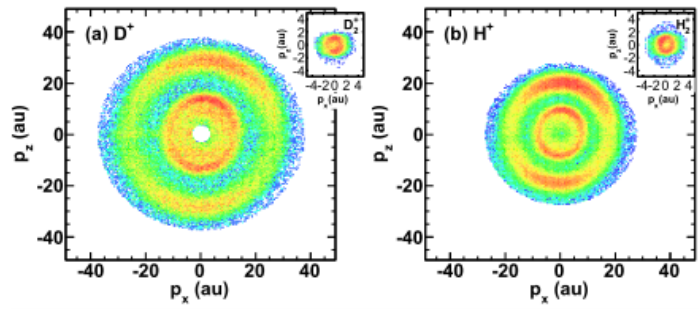

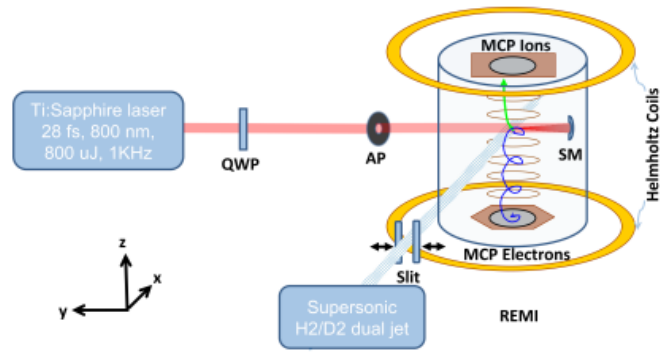

FIG. 1. (Color online). Momentum spectra for $D^+$ (a) and $H^+$ (b), as well as for $D_2^+$ and $H_2^+$ (respective insets). (c)The sketch of the experimental apparatus (REMI: reaction microscopes, AP: aperture, QWP: quarter-wave-plate).

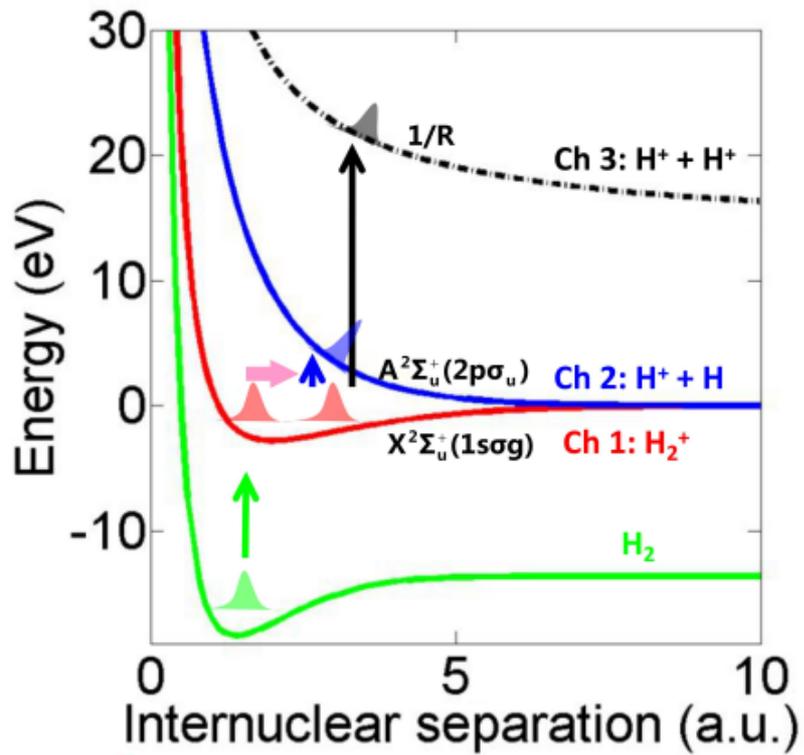

FIG. 2. (Color online). Potential energy curves of $H_2$ (green), two lowest electronic states $X^2\Sigma_g^+$ (red) and $A^2\Sigma_u^+$ (blue) of $H_2^+$, and $H_2^{++}$ Coulombic 1/R curve (R is internuclear distance). $H_2^+$ formed by single ionization can remain bound (Ch 1), dissociate (Ch 2) or undergo second ionization by the same pulse followed by Coulomb explosion (Ch 3).

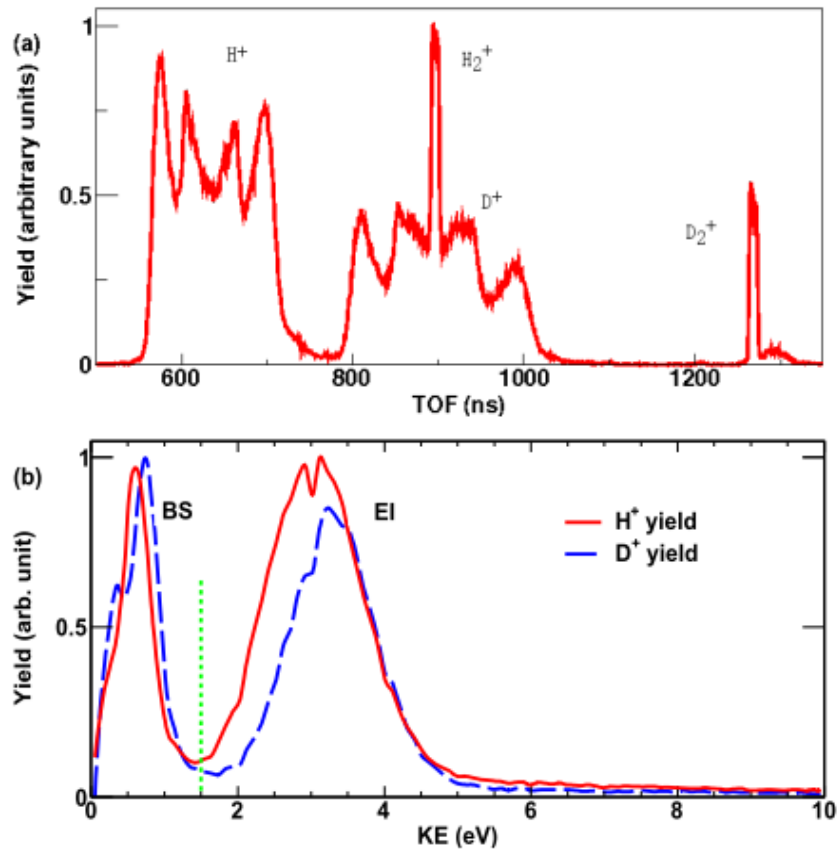

FIG. 3. (Color online). (a) The measured time-of-flight spectrum mixed $H_2/D_2$ gas ionized by circularly polarized pulses with centeral wavelegth of 800 nm, pulse duration FWHM of 28 fs and peak intensity of $(4.2 \pm 0.4) \times 10^{14}\ W/cm^2$. (b) The corresponding kinetic energy spectra for $H^+$ and $D^+$. The bond-softening (BS) and enhanced-ionization (EI) peaks are distinct in kinetic energy spectrum for both $H^+$ and $D^+$. Green dashed line represents the boundary between BS and EI bands used to determine the total overall single ionization yield.

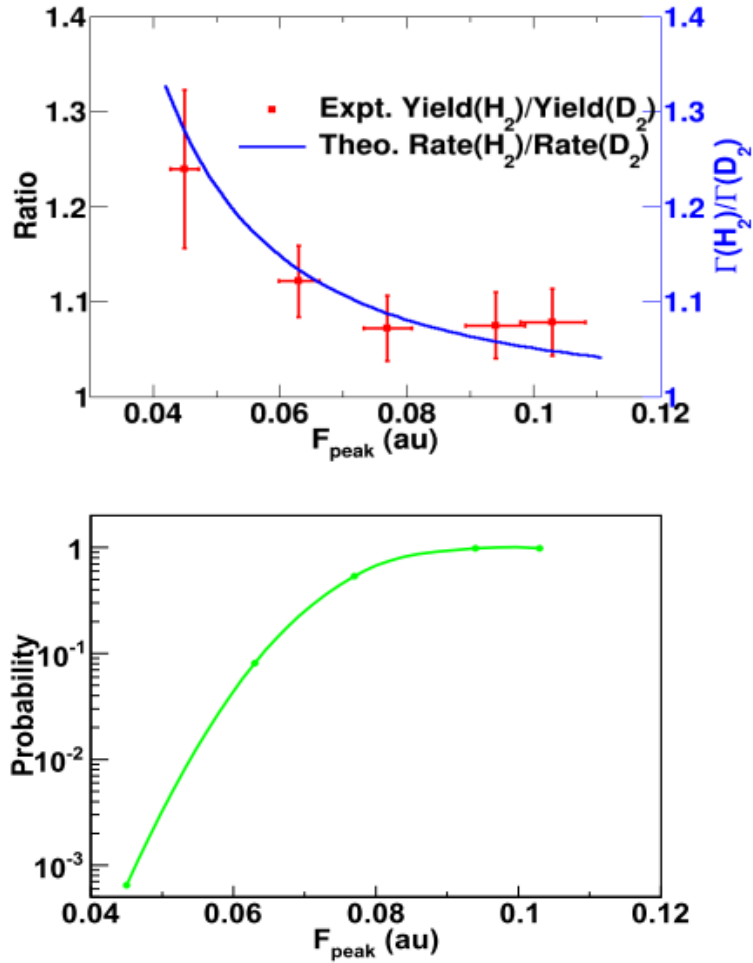

FIG. 4. (Color online). (a) The ratio of the total overall single ionization yield for $H_2$ and $D_2$ versus the laser peak field. The red squares are the experimental data. For comparison the ratio of the ionization rate for $H_2$ and $D_2$ predicted by the TWM theory [1] is plotted at the same time (blue solid line). (b) The calculated ionization probability vs. laser peak intensity for $H_2$.

# References:

∗ hanxu1981@gmail.com; † i.litvinyuk@griffith.edu.au